\documentclass[12pt]{article}
\usepackage{epsf,amsfonts,hyperref}
\usepackage{graphicx}

\renewcommand{\appendix}[1]{
    \addtocounter{section}{1}
    \setcounter{equation}{0}
    \renewcommand{\thesection}{\Alph{section}}
    \section*{Appendix \thesection\protect\indent #1}
    \addcontentsline{toc}{section}{Appendix \thesection\ \ \ #1}
}
\newcommand\encadremath[1]{\vbox{\hrule\hbox{\vrule\kern8pt
\vbox{\kern8pt \hbox{$\displaystyle #1$}\kern8pt}
\kern8pt\vrule}\hrule}}
\def\enca#1{\vbox{\hrule\hbox{
\vrule\kern8pt\vbox{\kern8pt \hbox{$\displaystyle #1$}
\kern8pt} \kern8pt\vrule}\hrule}}

\newcommand\figureframex[3]{
\begin{figure}[bth]
\hrule\hbox{\vrule\kern8pt
\vbox{\kern8pt \vbox{
\begin{center}
{\mbox{\epsfxsize=#1.truecm\epsfbox{#2}}}
\end{center}
\caption{#3}
}\kern8pt}
\kern8pt\vrule}\hrule
\end{figure}
}
\newcommand\figureframey[3]{
\begin{figure}[bth]
\hrule\hbox{\vrule\kern8pt
\vbox{\kern8pt \vbox{
\begin{center}
{\mbox{\epsfysize=#1.truecm\epsfbox{#2}}}
\end{center}
\caption{#3}
}\kern8pt}
\kern8pt\vrule}\hrule
\end{figure}
}

\renewcommand{\thesection}{\arabic{section}}

\makeatletter
\@addtoreset{equation}{section}
\makeatother
\newtheorem{theorem}{Theorem}[section]

\newtheorem{remark}{Remark}[section]
\newtheorem{proposition}{Proposition}[section]
\newtheorem{lemma}{Lemma}[section]
\newtheorem{corollary}{Corollary}[section]
\newtheorem{definition}{Definition}[section]
\def\br{\begin{remark}\rm\small}
\def\er{\end{remark}}
\def\bt{\begin{theorem}}
\def\et{\end{theorem}}
\def\bd{\begin{definition}}
\def\ed{\end{definition}}
\def\bp{\begin{proposition}}
\def\ep{\end{proposition}}
\def\bl{\begin{lemma}}
\def\el{\end{lemma}}
\def\bc{\begin{corollary}}
\def\ec{\end{corollary}}
\def\beaq{\begin{eqnarray}}
\def\eeaq{\end{eqnarray}}

\newcommand{\beq}{\begin{equation}}
\newcommand{\eeq}{\end{equation}}
\newcommand{\bea}{\begin{eqnarray}}
\newcommand{\eea}{\end{eqnarray}}

\renewcommand{\and}{{\qquad {\rm and} \qquad}}

\newcommand{\Res}{\mathop{\,\rm Res\,}}

\newcommand{\td}[1]{{\tilde{#1}}}

\newcommand{\Pint}{{\int\kern -1.em -\kern-.25em}}

\newcommand{\acycle}{{\cal A}}
\newcommand{\bcycle}{{\cal B}}

\newcommand{\spcurve}{{\cal S}}
\newcommand{\curve}{{\cal C}}

\newcommand{\genus}{{\mathfrak g}}
\newcommand{\CP}{{\mathbb CP^1}}

\begin{document}
%=============================Page de titre==============%\date{??}
%\author{Eynard}
%\title{Correlation functions for hermitian random matrices}
%\topmargin .5cm \textheight 21.5cm \textwidth 15.8cm
%\oddsidemargin 0.54cm
%\evensidemargin 0.54cm
\sloppy

%\maketitle

\pagestyle{empty}
\hfill IPhT-T13/252, CRM-3330
\addtolength{\baselineskip}{0.20\baselineskip}
\begin{center}
\vspace{26pt}
{\large \bf {About the x-y symmetry of the $F_g$
algebraic invariants.}}
\newline
\vspace{26pt}

{\sl B.\ Eynard}\footnote{bertrand.eynard@cea.fr}${}^{a,b}$,\hspace*{0.05cm}{\sl N.\ Orantin}\footnote{norantin@math.ist.utl.pt}\hspace*{0.05cm}\\ 
\vspace{12pt}
${}^a$ Institut de Physique Th\'eorique,
CEA, IPhT, F-91191 Gif-sur-Yvette, France CNRS, URA 2306, F-91191 Gif-sur-Yvette, France\\
${}^b$ CRM, Centre de Recherche Math\'ematiques, Montr\'eal QC Canada.\\
\end{center}

\vspace{20pt}
\begin{center}
{\bf Abstract}:
We complete the proof of the $x-y$ symmetry of symplectic invariants of \cite{EOFg}. We recall the main steps of the proof of \cite{EOFgsymxy}, and we include the integration constants absent in \cite{EOFgsymxy}.
\end{center}

%-----------------------------ABSTRACT--------------------------------------
%
%Abstract

%\begin{center}

%\end{center}

%\newpage
%\pagestyle{empty}

%\section*{}

%\newpage
\vspace{26pt}
\pagestyle{plain}
\setcounter{page}{1}

%*********************************************************************
%==================== ARTICLE =======================================%*********************************************************************

\newpage

%\tableofcontents

\section{Introduction}

In \cite{EOFg} we introduced some "invariants" $F_g(\spcurve)$ associated to an algebraic curve $\spcurve=\{(x,y)\,|\,P(x,y)=0\}$ immersed in $\mathbb C\times\mathbb C$.
We claimed in \cite{EOFgsymxy} that the $F_g$'s were invariant under the symmetry $x\leftrightarrow y$ , i.e. if $\td\spcurve=\{(y,x)\,|\,P(x,y)=0\}\subset \mathbb C\times\mathbb C$, we had $F_g(\spcurve)=F_g(\td\spcurve)$.
In fact, the identity $F_g(\spcurve)=F_g(\td\spcurve)$ holds only for certain classes of spectral curves $\spcurve$, typically those appearing in the 2-matrix model as in \cite{EOFgsymxy}. However, as noticed by V. Bouchard and P. Sulkowski in \cite{BS} and later developed in \cite{BHLMR}, it can be wrong for more general classes of curves, because of some integration constants which were disregarded in \cite{EOFgsymxy}.
The actual invariance, valid for any algebraic spectral curve is:
\beq
\hat F_g(\spcurve) = F_g(\spcurve) + \frac{1}{2-2g} \sum_i t_i \int_*^{\alpha_i} \omega_{g,1}(\spcurve),
\eeq
for which we have:
\beq
\hat F_g(\spcurve) = \hat F_g(\td\spcurve).
\eeq

In this article we recall the main steps of \cite{EOFgsymxy}, and we include the integration constants missing in \cite{EOFgsymxy} in order to prove the symmetry property for the corrected $\hat F_g$.

\section{Spectral curves and their invariants}

\bd
An algebraic  spectral curve $\spcurve=(\curve,x,y)$ is the data of a compact Riemann surface $\curve$ of genus $\genus$, together with a choice of $2\genus$ independent non-contractible cycles $\acycle_1,\dots,\acycle_\genus,\bcycle_1,\dots,\bcycle_\genus$ on it with symplectic intersections:
\beq
\acycle_i\cap\bcycle_j=\delta_{i,j}\, , \quad
\acycle_i\cap\acycle_j=0\, , \quad
\bcycle_i\cap\bcycle_j=0\, , \quad
\eeq
and $x$ and $y$ are two meromorphic functions $\curve\to \CP$.

Moreover, we say that $\spcurve=(\curve,x,y)$ is a regular spectral curve, if $dx$ has only simple zeroes on $\curve$, and the zeroes of $dx$ are distinct from the zeroes of $dy$ and from the poles of $x$ and $y$.

\ed

The map $\curve \to \mathbb C\times \mathbb C, z\mapsto (x(z),y(z))$ defines an algebraic curve immersed in $\mathbb C\times \mathbb C$. 

\bd
The zeroes of $dx$ are called the "branchpoints":
\beq
\mathbf a = \{a_1,\dots, a_s\} \quad ,  \, dx(a_i)=0.
\eeq
In a vicinity $U_a$ of a branchpoint $a$, a good local coordinate is $\zeta_a(z)=\sqrt{(x(z)-x(a))}$.

The local Galois involution $s_a:U_a\to U_a$ is such that $x\circ s_a=x$, and $s_a\neq {\rm Id}$.
In the local coordinate $\zeta_a(z)=\sqrt{(x(z)-x(a))}$, the local Galois involution is simply:
\beq
s_a\left(\sqrt{(x(z)-x(a))}\right) = - \,\sqrt{(x(z)-x(a))}.
\eeq

\ed

Following \cite{EOFg}, to a regular spectral curve $\spcurve=(\curve,x,y)$ we associate its invariants:
\bd
The invariants $\omega_{g,n}(\spcurve)$ are symmetric meromorphic differentials $\in K(\curve)^{\otimes n}$ (where $K(\curve)$ is the canonical bundle of $\curve$), such that:

$\bullet \qquad \qquad \omega_{0,1}=ydx$,

$\bullet$ $\omega_{0,2}$ is the fundamental second kind form on $\curve$ \cite{mumford}, i.e. the  unique bilinear differential on $\curve\times \curve$, with a normalized double pole on the diagonal, and no other pole:
\beq
\omega_{0,2}(z_1,z_2) \mathop{\sim}_{z_2\to z_1} \frac{dz_1\,dz_2}{(z_1-z_2)^2} + {\rm analytical}
\eeq
and normalized on the $\acycle$-cycles:
\beq
\forall\, z_1 \in \curve \, , \qquad \oint_{z_2\in\acycle_i} \omega_{0,2}(z_1,z_2) = 0;
\eeq

$\bullet$ for $n\geq 1$ and $(g,n)\neq (0,1),(0,2)$, the $\omega_{g,n}(\spcurve)$ are computed by the topological recursion of \cite{EOFg}:
\bea
\omega_{g,n}(z_1,\overbrace{z_2,\dots,z_n}^{J})
&=& \sum_{a\in\mathbf a} \Res_{z\to a} K_a(z_1,z)\,\Big(\omega_{g-1,n+1}(z,s_a(z),J) \cr
&& + \sum'_{h+h'=g,\, I\cup I'=J} \omega_{h,1+|I|}(z,I)\,\omega_{h',1+|I'|}(s_a(z),I')\Big)
\eea
where the recursion kernel is
\beq
K_a(z_1,z) = -\frac{1}{2}\,\, \frac{\int_{z'=s_a(z)}^z \omega_{0,2}(z_1,z')}{\omega_{0,1}(z)-\omega_{0,1}(s_a(z))}
\eeq
and the $\sum'$ means that we exclude the terms $(h,I)=(0,\emptyset)$ and $(h,I)=(g,J)$.

\medskip

The scalar invariants  $\omega_{g,0}(\spcurve)=F_g(\spcurve)\in \mathbb C$, are given by:
\beq
\forall\,g\geq 2\,, \qquad
F_g = \frac{1}{2-2g}\,\sum_{a\in\mathbf a} \Res_{z\to a} \omega_{g,1}(z) \, \Phi(z)
\eeq
where $d\Phi=\omega_{0,1}=ydx$ in a vicinity of each ${a\in\mathbf a}$.

\ed

\br
We shall not consider $F_0$ and $F_1$ in this article, their $x-y$ symmetry properties have already been established.
\er

\section{The x-y symmetry}

Now, consider the two spectral curves 
\beq
\spcurve=(\curve,x,y)
\qquad , \qquad 
\td\spcurve=(\curve,y,x)
\eeq
with the same compact Riemann surface $\curve$ and the same choice of independent contours $\left( {\cal A}_i,{\cal B}_i \right)_{i=1,\dots,\genus}$, and
which we assume are both regular.
Let:
\beq
\mathbf a = \{a_1,\dots,a_s\} = {\rm zeroes\,of}\,\,dx
\quad , \quad
\mathbf b = \{b_1,\dots,b_{\td s}\} = {\rm zeroes\,of}\,\,dy
\eeq
We shall need to consider the poles of $x$ and $y$, we call them:
\beq
\mathbf\alpha = \{\alpha_1,\dots,\alpha_r\} = {\rm poles\,of}\,\, x\,{\rm and}\,y.
\eeq
\beq
d_i=\deg_{\alpha_i} x
\qquad , \qquad
\td d_i=\deg_{\alpha_i} y.
\eeq

We shall define the times:
\beq\label{deftimes}
t_i = \Res_{z\to \alpha_i} ydx  = -\td t_i = - \Res_{z\to \alpha_i} xdy.
\eeq

We shall denote:
\beq
\omega_{g,n}\equiv \omega_{g,n}(\spcurve)
\quad , \quad 
\td\omega_{g,n}\equiv \omega_{g,n}(\td\spcurve).
\eeq
and
\beq
F_g=F_g(\spcurve) \quad , \quad \td F_g=F_g(\td\spcurve).
\eeq

Our goal is to compare the invariants, i.e. compute
\beq
F_g-\td F_g  = ?\,
\eeq

\subsection{Sketch of the construction of \cite{EOFgsymxy}}

The main idea in \cite{EOFgsymxy} is to define by a recursion a sequence of differentials for any $g\geq 0$ and any $n+m>0$:
\beq
\omega_{g,n,m}(\spcurve)=\omega_{g,m,n}(\td\spcurve) \in K(\curve)^{\otimes m+n},
\eeq
which are by construction manifestly symmetric in the exchange of $x$ and $y$.

(These definitions in \cite{EOFgsymxy} may look complicated, but they are simply obtained by mimicking the loop equations in a 2 matrix model).

It was proved in \cite{EOFgsymxy} that 
\bp[\cite{EOFgsymxy}]
The differential forms $\omega_{g,n,m}$ satisfy:

$\bullet$ For any $n \geq 1$,
\beq
\omega_{g,n,0}=\omega_{g,n}
\qquad , \qquad 
\omega_{g,0,n}=\td\omega_{g,n} ;
\eeq
In particular $\omega_{0,1,0}=ydx$, $\omega_{0,0,1}=xdy$, $\omega_{0,2,0}=\omega_{0,0,2}=$fundamental second kind differential.

$\bullet$ if $2g-2+n+m>0$, $\omega_{g,n,m}(z_1,\dots,z_n;\td z_1,\dots,\td z_m)$ has poles only when $z_i\in\mathbf a$, $\td z_j\in\mathbf b$, and whenever $x(z_i)=x(\td z_j)$ or $y(z_i)=y(\td z_j)$.

$\bullet$ Let $\mathbf q=\{q_1,\dots,q_n\}$ and $\mathbf p=\{p_1,\dots,p_m\}$, the following form is exact (with respect to the variable $z\in \curve$) :
\beq
\omega_{g,n+1,m}(z,\mathbf q;\mathbf p) + \omega_{g,n,m+1}(\mathbf q;z,\mathbf p) 
= d_z\left( \frac{A_{g,n,m}(z;\mathbf q;\mathbf p)}{dx(z)\,dy(z)}\right)
\eeq
where $A_{g,n,m}(z;\mathbf q;\mathbf p)$ is a quadratic differential of $z\in\curve$, which has poles at $z\in\mathbf a$ and $z\in \mathbf b$ and when $x(z)=x(q_i)$ or $y(z)=y(q_i)$ or $x(z)=x(p_j)$ or $y(z)=y(p_j)$.
It may also have poles at the poles $z\to \alpha_i$;

$\bullet$ $d_z(A_{g,n,m}(z;\mathbf q;\mathbf p)/dx(z)dy(z))$ vanishes to order $d_i+\td d_i$ at  a pole $z=\alpha_i$.

\ep

In particular, if $n=m=0$ we have:
\beq
\omega_{g,1}(z)+\td\omega_{g,1}(z)=\omega_{g,1,0}(z) + \omega_{g,0,1}(z) = d_z\,\left( \frac{A_{g,0,0}(z)}{dx(z)\,dy(z)}\right)
\eeq
where $A_{g,0,0}(z)$ is a quadratic differential on $\curve$, whose only poles are¤ at $z\in\mathbf a\cup \mathbf b$.
$d_z(A_{g,0,0}(z)/dx(z)dy(z))$ vanishes to order $d_i+\td d_i$ at  a pole $z=\alpha_i$.

This implies, that,  in the vicinity of $\alpha_i$, there exists a choice of integration constant $C_{g;i}$, such that:
\beq
A_{g,0,0}(z) - C_{g;i}\,dx(z)\,dy(z) = D_{g,i}(z) \, dx(z) \, dy(z)
\eeq
where $D_{g,i}(z)$  vanishes to order $d_i+\td d_i+1$ at $\alpha_i$.

The integration constants satisfy
\beq
C_{g;i} - C_{g;j} = \int_{\alpha_j}^{\alpha_i} \left[ \omega_{g,1}(z)+\td\omega_{g,1}(z) \right].
\eeq

Since the sum of residues of a 1-form must be zero, we have:
\beq
\sum_i t_i = \sum_i \Res_{\alpha_i} ydx = 0 .
\eeq
Therefore
\beq\label{defCgi}
\sum_i t_i C_{g;i} = \sum_i t_i\,\int_{o}^{\alpha_i} \left[\omega_{g,1}(z)+\td\omega_{g,1}(z)\right]
\eeq
is independent of a choice of origin $o\in\curve$.

\subsection{Symmetry of the $F_g$'s}

Let us define $\Phi$ and $\td\Phi=xy-\Phi$ as functions on some vicinity of the branchpoints, such that:
\beq
d\Phi=ydx
\qquad , \qquad d\td\Phi = xdy.
\eeq
We have, by definition of the $F_g$'s  for $g\geq 2$:
\beq
F_g = \frac{1}{2-2g}\,\sum_{a \in \mathbf a} \Res_{z\to a} \omega_{g,1}(z)\Phi(z)
\quad , \quad 
\td F_g = \frac{1}{2-2g}\,\sum_{b \in \mathbf b} \Res_{z\to b} \td\omega_{g,1}(z)\td\Phi(z) .
\eeq
This implies that
\beq
(2-2g)\,(F_g-\td F_g)
=\sum_{a \in \mathbf a} \Res_{z\to a} \omega_{g,1}(z)\Phi(z) - \sum_{b \in \mathbf b} \Res_{z\to b} \td\omega_{g,1}(z)\td\Phi(z).
\eeq
Notice that, since $\omega_{g,1}$ and $\Phi$ have no pole at the $b_i$'s (zeroes of $dy$) and  $\td\omega_{g,1}$ and $\td\Phi$ have no pole at the $a_i$'s (zeroes of $dx$), we may write:
\bea
(2-2g)\,(F_g-\td F_g)
&=& \sum_{a\in \mathbf a\cup\mathbf b} \Res_{z\to a} \omega_{g,1}(z)\Phi(z) -  \td\omega_{g,1}(z)\td\Phi(z) \cr
&=& \sum_{a\in \mathbf a\cup\mathbf b} \Res_{z\to a} \omega_{g,1}(z)\Phi(z) -  \td\omega_{g,1}(z)(x(z)y(z)-\Phi(z)) \cr
&=& \sum_{a\in \mathbf a\cup\mathbf b} \Res_{z\to a} (\omega_{g,1}(z)+  \td\omega_{g,1}(z))\Phi(z)  - \td\omega_{g,1}(z)\,x(z)y(z) .\cr
\eea
It was proved in \cite{EOFg} that, for any spectral curve,
\beq
\Res_{z\to a} \omega_{g,1}(z)\,x(z)\,y(z) =0.
\eeq
Therefore we have:
\bea
(2-2g)\,(F_g-\td F_g)
&=& \sum_{a\in \mathbf a\cup\mathbf b} \Res_{z\to a} (\omega_{g,1}(z)+  \td\omega_{g,1}(z))\Phi(z)  \cr
&=& \sum_{a\in \mathbf a\cup\mathbf b} \Res_{z\to a} \Phi(z)\,d\left(\frac{A_{g,0,0}(z)}{dx(z)\,dy(z)}\right)  \cr
\eea
and, by integrating by parts,
\bea
(2-2g)\,(F_g-\td F_g)
&=& - \sum_{a\in \mathbf a\cup\mathbf b} \Res_{z\to a} \frac{A_{g,0,0}(z)}{dx(z)\,dy(z)}\,\,d\Phi(z)  \cr
&=& - \sum_{a\in \mathbf a\cup\mathbf b} \Res_{z\to a} \frac{A_{g,0,0}(z)}{dx(z)\,dy(z)}\,\,y(z)\,dx(z)  \cr
&=& - \sum_{a\in \mathbf a\cup\mathbf b} \Res_{z\to a} A_{g,0,0}(z)\,\,\frac{y(z)}{dy(z)}  .\cr
\eea
Now, let us move the integration contour, so that we enclose all the other poles of $A_{g,0,0}$, i.e. the $\alpha_i$'s.
We have:
\bea
(2-2g)\,(F_g-\td F_g)
&=&  \sum_{i} \Res_{z\to \alpha_i} A_{g,0,0}(z)\,\,\frac{y(z)}{dy(z)}  \cr
&=&  \sum_{i} \Res_{z\to \alpha_i} (C_{g;i}dx(z)\,dy(z)+D_{g,i}(z) \, dx(z) \, dy(z))\,\,\frac{y(z)}{dy(z)} . 
\eea
Since $D_{g,i}(z)$ vanishes to order $d_i+ \tilde{d}_i+1$ while  $y(z) dx(z)$ has a pole of order $d_i+ \tilde{d}_i+1$, the second term $D_{g,i}(z) y(z) dx(z)$ is regular at the pole $\alpha_i$, so that:
\bea
(2-2g)\,(F_g-\td F_g)
&=&  \sum_{i} C_{g;i}\,\Res_{z\to \alpha_i} y(z)\,dx(z)  \cr
&=&  \sum_{i} t_i\,C_{g;i}\,  
\eea
and, according to eq.\ref{defCgi},
\beq
\sum_i t_i C_{g;i} = \sum_i t_i\,\int_{o}^{\alpha_i} \left[ \omega_{g,1}(z)+\td\omega_{g,1}(z)\right].
\eeq
We find
\bea
(2-2g)\,(F_g-\td F_g)
&=&  \sum_i t_i\,\int_{o}^{\alpha_i} \left[ \omega_{g,1}(z)+\td\omega_{g,1}(z) \right]  \cr
&=&  \sum_i t_i\,\int_{o}^{\alpha_i} \omega_{g,1}(z) - \sum_i \td t_i\,\int_{o}^{\alpha_i} \td\omega_{g,1}(z) . \cr
\eea
This implies that
\beq
F_g - \frac{1}{2-2g}\sum_i t_i\,\int_{o}^{\alpha_i} \omega_{g,1} 
=
\td F_g - \frac{1}{2-2g}\sum_i \td t_i\,\int_{o}^{\alpha_i} \td\omega_{g,1} 
\eeq
and thus:

\bt
The following quantity:
\beq
\hat F_g(\spcurve) = F_g(\spcurve) - \frac{1}{2-2g}\sum_i \left(\Res_{\alpha_i} \omega_{0,1}(\spcurve)\right)\,\,\left(\int_{o}^{\alpha_i} \omega_{g,1}(\spcurve) \right) 
\eeq
(which is independent of a choice of a generic basepoint $o\in\curve$)
is invariant under the exchange $(x,y)\leftrightarrow (y,x)$:
\beq
\hat F_g(\td\spcurve) = \hat F_g(\spcurve).
\eeq
\et

\section{Conclusion}

We have completed the proof of the $(x\leftrightarrow y)$ symmetry of \cite{EOFgsymxy}, by including the integration constants.
We see that $\hat F_g=F_g+{\rm integration\,constants}$, is symplectic invariant, rather than $F_g$.

\medskip
Remark that in the context of the 2-matrix model, and their scaling limit which is the $(p,q)$ minimal models for which $t_i=\Res ydx =0$, the integration constants were absent, and thus the $F_g$'s were indeed invariant.
This proves that
\beq
F_g((p,q)\,{\rm minimal\,model}) = F_g((q,p)\,{\rm minimal\,model}).
\eeq

\section*{Aknowledgements}

This work was motivated by the remarks of Vincent Bouchard, who did check numerically (in particular in \cite{BS} with Piotr Sulkowski and in \cite{BHLMR} with J. Hutchinson, P. Loliencar, M. Meiers and M. Rupert) the $(x\leftrightarrow y)$ symmetry on many examples, and empirically observed that the $\hat F_g$'s were invariant. After his remarks we quickly found the missing integration constants $C_{g,i}$ from \cite{EOFgsymxy}, and here is the correction. We are grateful to him.
The work of B.E. is supported by the Quebec government by the FQRNT fund, and by the ERC starting grant Field-knots with P. Sulkowski.

\end{document}